# Mid-Infrared Plasmonic Biosensing with Graphene

Daniel Rodrigo,[1] Odeta Limaj,[1] Davide Janner,[2] Dordaneh Etezadi,[1]
F. Javier García de Abajo,[2,3] Valerio Pruneri,[2,3] Hatice Altug[1]*

**Affiliations:**

[1] Institute of BioEngineering, École Polytechnique Fédérale de Lausanne (EPFL), CH-1015 Lausanne, Switzerland.

[2] ICFO - Institut de Ciències Fotòniques, Mediterranean Technology Park, 08860 Castelldefels (Barcelona), Spain

[3] ICREA - Institució Catalana de Recerca i Estudis Avançats, 08010 Barcelona, Spain

* Corresponding author. E-mail: hatice.altug@epfl.ch

**Abstract**:

Infrared spectroscopy is the technique of choice for chemical identification of biomolecules through their vibrational fingerprints. However, infrared light interacts poorly with nanometric size molecules. Here, we exploit the unique electro-optical properties of graphene to demonstrate a high-sensitivity tunable plasmonic biosensor for chemically-specific label-free detection of protein monolayers. The plasmon resonance of nanostructured graphene is dynamically tuned to selectively probe the protein at different frequencies and extract its complex refractive index. Additionally, the extreme spatial light confinement in graphene, up to two orders of magnitude higher than in metals, produces an unprecedentedly high overlap with nanometric biomolecules, enabling superior sensitivity in the detection of their refractive index and vibrational fingerprints. The combination of tunable spectral selectivity and enhanced sensitivity of graphene opens exciting prospects for biosensing.

**Main Text:**

Graphene has the potential to reshape the landscape of photonics and optoelectronics owing to its exceptional optical and electrical properties (*1-3*). In particular, its infrared (IR) response is characterized by long-lived collective electron oscillations (plasmons) that can be dynamically tuned by electrostatic gating, in contrast to conventional plasmonic materials such as noble metals (*4-10*). Furthermore, the electromagnetic fields of graphene IR plasmons display unprecedented spatial confinement, making them extremely attractive for enhanced light-matter interactions and integrated mid-IR photonics (*11-14*). Specifically, biosensing is an area in which graphene tunability and IR light localization offer great opportunities.

The mid-IR range is particularly well-suited for biosensing as it encompasses the molecular vibrations that uniquely identify the biochemical building blocks of life, such as proteins, lipids, and DNA (*15*). IR absorption spectroscopy is a powerful technique that provides exquisite biochemical information in a non-destructive label-free fashion by accessing these vibrational fingerprints. Nevertheless, vibrational absorption signals are prohibitively weak due to the large mismatch between mid-IR wavelengths (2-6µm) and biomolecular dimensions (<10nm). To overcome this limitation, high sensitivity can be achieved by exploiting the strong optical near fields in the vicinity of resonant metallic nanostructures (*16-18*), which comes at the expense of a reduced spectral bandwidth and is ultimately limited by the relatively poor field confinement of metals in the mid-IR (*19*).

In this work we report a graphene-based tunable mid-IR biosensor and demonstrate its potential for quantitative protein detection and chemical-specific molecular identification. Our device (Fig.1A) consists of a CVD graphene layer deposited on a 280nm-thick native silica oxide of a silicon substrate. Graphene nanoribbon arrays (width $W$=20-60nm and period $P$~$2W$) are then patterned using e-beam lithography and oxygen plasma etching (*20*). Scanning electron microscope (SEM) and atomic force microscope (AFM) cross-section for typical samples are shown in Figs. 1B and 1C. We apply an electrostatic field across the $SiO_2$ layer through a bias voltage ($V_g$) varied in the 0-120V range to dynamically control the Fermi level ($E_F$) of graphene. Extinction spectra of the device are acquired using Fourier Transform Infrared (FTIR) spectroscopy for incident electric field polarized perpendicular to the nanoribbons. Figure 2A shows the extinction for a nanoribbon array with $W$=30nm, $P$=80nm, and different $V_g$ (dashed curves). A prominent resonance is observed which is associated with localized surface plasmons (LSP) polarized across the nanoribbons. By changing $V_g$, the resonance frequency is tuned continuously from 1450 to above 1800cm$^{-1}$. The ribbon width $W$=30nm is chosen so that the frequency tuning range sweeps across the target vibrational fingerprints (Fig. S1).

We aim at detecting protein molecules, the primary material of life enabling most of the critical biological functions. The main vibrational fingerprints of proteins are amide I and II bands (1660 and 1550cm$^{-1}$), which are associated with the C=O stretch and N-H wag / C-N modes in the amide functional group. For demonstration of protein detection we use recombinant protein A/G and goat anti-mouse IgG antibody. Incubation of A/G on the sensor surface allows the formation of a protein monolayer by physisorbtion, which is then used to bind IgG antibodies and form a well-defined protein bilayer (*20*). The extinction spectra of the sensor are presented in Fig. 2A before and after protein bilayer formation showing dramatic changes upon protein immobilization. The first observed prominent effect is a redshift of the plasmonic resonance as a consequence of the change in the refractive index at the sensor surface. Despite the nanometric thickness of the protein bilayer, we detect frequency shifts exceeding 200cm$^{-1}$. The second prominent effect is the emergence of two spectral dips at 1660cm$^{-1}$ and 1550cm$^{-1}$ that are almost undetectable when they are far from the plasmonic resonance (e.g., for $V_g$=-20 V) and become progressively more intense with increasing spectral overlap (e.g., for $V_g$=-130 V). Their spectral positions coincide with amide I and II bands, respectively, unambiguously revealing the presence of the protein compounds in a chemically-specific manner. The decrease in extinction induced by the vibrational modes is the result of resonant coupling between plasmons and molecular vibrations (*21*).

In order to extract quantitative information on the protein optical parameters, we use an analytical model of the IR response of the graphene nanoribbon array (*22*). We model graphene in the electrostatic limit ($W, P \ll \lambda$) and assume that the ribbon response is dominated by the lowest-order transversal mode. The transmission coefficient of the structure then reduces to

$$t = t_0 + \frac{4\pi^2 i \ \alpha^{\text{eff}}(\omega) \ (1 + r_0) \ t_0}{n_2 \ P \ \lambda}$$

, where

$$\alpha^{\text{eff}}(\omega) = \frac{0.894 \ W^2}{\frac{A}{n_1^2 + n_2^2} - \frac{i \ \omega W}{\sigma(\omega)}}$$

is an effective graphene-ribbon polarizability that takes into account the complex refractive indices of the silica substrate $n_2$ (*23*) and the material immediately above the ribbons $n_1$, while the coefficient A is a function of $P/W$ (in particular, A=28.0 for the $P/W$=2.67). Here, $t_0$ and $r_0$ are the transmission and reflection coefficients of the interface between media 1 and 2 in the absence of graphene. The response of the latter enters through its frequency-dependent surface conductivity $\sigma(\omega)$, which we model in the local-RPA approximation (*11*). Finally, we compute the ratio of transmission in regions with and without graphene as $|t/t_0|^2$, which is the magnitude measured in the experiments.

The analytic model is first used to extract the graphene parameters from experimental IR spectra for bare nanoribbons (i.e., with $n_1$=1). The calculated spectra are reported in Fig. 2B (dashed curves) for the extracted relaxation time ($\tau$=15fs) and Fermi energies ($E_F$=0.18 to 0.43eV). We observe that the carrier density ($n_s \sim E_F^2$) changes linearly with $V_g$ (Fig. 2C) and has an intrinsic doping $E_{F0}$=0.17eV produced by charge transfer from the silica. Next, the analytic model is used to retrieve the protein permittivity from experimental results by adjusting a Lorentzian permittivity $\varepsilon_p = n_1^2 = n_\infty^2 + \sum S_k^2/(\omega_k^2 - \omega^2 - i\omega\gamma_k)$. Good agreement is observed between experimental and calculated spectra (Fig.2B) for the protein Lorentzian parameters upon least-square fitting. The extracted permittivity has a non-dispersive term $n_\infty^2$=2.08 and shows two absorption peaks at 1668 and 1532cm$^{-1}$, matching amide I and II bands respectively (Fig. 2D). The fitted permittivity is also in good agreement with independent protein permittivity measurements from ellipsometry ($n_\infty^2$) and IR reflection absorption spectroscopy IRRAS ($S_k, \omega_k, \gamma_k$) (20). There is however a small discrepancy, which we attribute to a slight overestimate of plasmon-protein coupling in the theoretical model. These results indicate that the proposed graphene biosensor combines refractive index sensing, so far a prerogative of visible plasmonic sensors, with the unique chemical specificity of mid-IR spectroscopy, together with the extra degree of freedom enabled by the graphene electro-optical tunability.

The characteristics of our graphene biosensor become more evident by comparing its spectral response to that of a state-of-the-art metallic localized surface plasmon resonance (LSPR) sensor composed of a gold dipole-antenna array (Fig. 3). Both devices are first operated in a spectral range free of protein vibrational modes by setting graphene at $Vg$=-20V and designing a gold dipole length $L$=2.6µm (Fig. 3A). Upon protein immobilization, we detect a resonance shift of 160cm$^{-1}$ for graphene, which is approximately 6 times larger than the 27cm$^{-1}$ shift obtained with gold. Next, the operation spectrum is moved towards the protein amide I and II bands by setting graphene at $Vg$=-120V and using a different gold sensor with $L$=2.1µm (Fig. 3B). Clearly, dynamic tunability of graphene is one of its main advantages over gold for surface enhanced IR absorption (SEIRA), enabling sensing over a broad spectrum with a single device. In addition, for the SEIRA signal corresponding to amide I band the graphene sensor features a signal modulation of 27%, which is almost 3 times larger than the one observed with the gold sensor (11%).

The large spectral shifts and absorption signals confirm the unprecedented sensitivity of our graphene biosensor to the complex refractive index of the target molecule. For similar IR frequency plasmons, the graphene atomic thickness leads to a higher confinement, resulting in a much larger spatial overlap between the mid-IR plasmonic field and the analyte. Figure 3C shows the near-field distribution of LSPR modes in graphene nanoribbons and gold dipole arrays calculated with a finite element method. The field hotspots are located at the end-points of the gold dipole and along the edges of the graphene nanoribbon. We compute the percentage of near field intensity confined within a given distance $d$ from the structure (Fig. 3D). We observe that 90% of the mode energy is confined within 15 nm from the graphene surface, while the same percentage is spread over a distance of 500 nm away from the gold surface, thus confirming the tighter field confinement of graphene in the mid-IR. As the biosensing signal comes only from the field inside the target volume, we also calculate the field overlap with an 8-nm-thick protein bilayer, which is 29% for graphene, while it is only 4% for gold. The near-field intensity overlap can be experimentally extracted as the ratio of the relative resonance shift ($\Delta\omega/\omega$) and the permittivity variation ($\varepsilon_{protein}$-1) (24). This estimate yields 26% and 5% field overlap for graphene and gold, in good agreement with simulations (see above). These results demonstrate the ability of graphene to provide stronger light-protein interactions beyond state-of-the-art metallic plasmonic sensors, and further improvement in the graphene quality should lead to even better sensitivity and spectral resolution.

**Acknowledgments:** This work was funded in part by European Commission (FP7-IEF-2013-625673-GRYPHON, Graphene Flagship CNECT-ICT-604391 and FP7-ICT-2013-613024-GRASP), by Spanish Ministry of Economy and Competitiveness (MINECO), "Fondo Europeo de Desarrollo Regional" (FEDER) through grant TEC2013-46168-R, by NATO's Public Diplomacy Division in the framework of "Science for Peace", by European Union's Horizon 2020 research and innovation programme under grant agreement No 644956, by Fundació Privada Cellex, the Severo Ochoa Program and Ramon y Cajal fellowship Program. We also acknowledge École Polytechnique Fédérale de Lausanne and Center of MicroNano Technology for finantial support and nanofabrication.

**Fig. 1.**

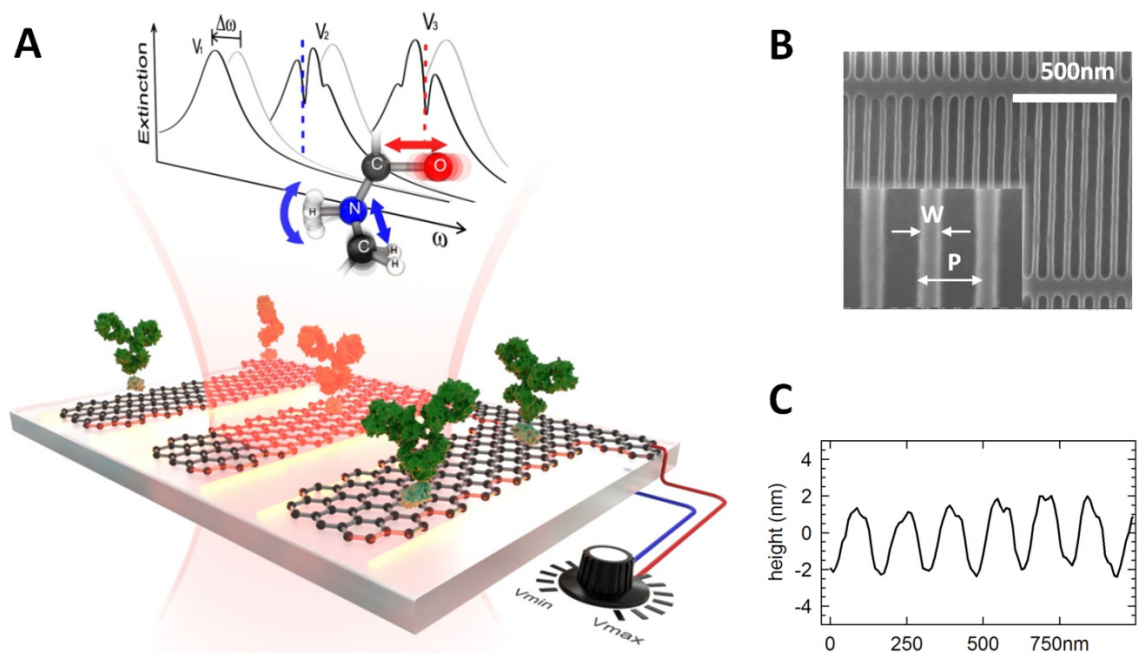

**Fig. 1. Tunable graphene mid-IR biosensor.** (**A**) Conceptual view of the graphene biosensor. An infrared beam excites a plasmon resonance across the graphene nanoribbons. The electromagnetic field is concentrated at the ribbon edge, enhancing light interaction with the protein molecules adsorbed by graphene. Protein sensing is achieved by detecting a plasmon resonance spectral shift ($\Delta\omega$) accompanied by narrow dips corresponding to the molecular vibration bands of the protein. The plasmonic resonance is electrostatically tuned to sweep continuously over the protein vibrational bands. (**B**) Scanning electron microscope image of a graphene nanoribbon array (width W=30nm, period P=80nm). Vertical nanoribbons are electrically interconnected by horizontal strips to maintain the graphene surface at uniform potential. (**C**) Atomic force microscope cross-section of a graphene nanoribbon array.

**Fig. 2.**

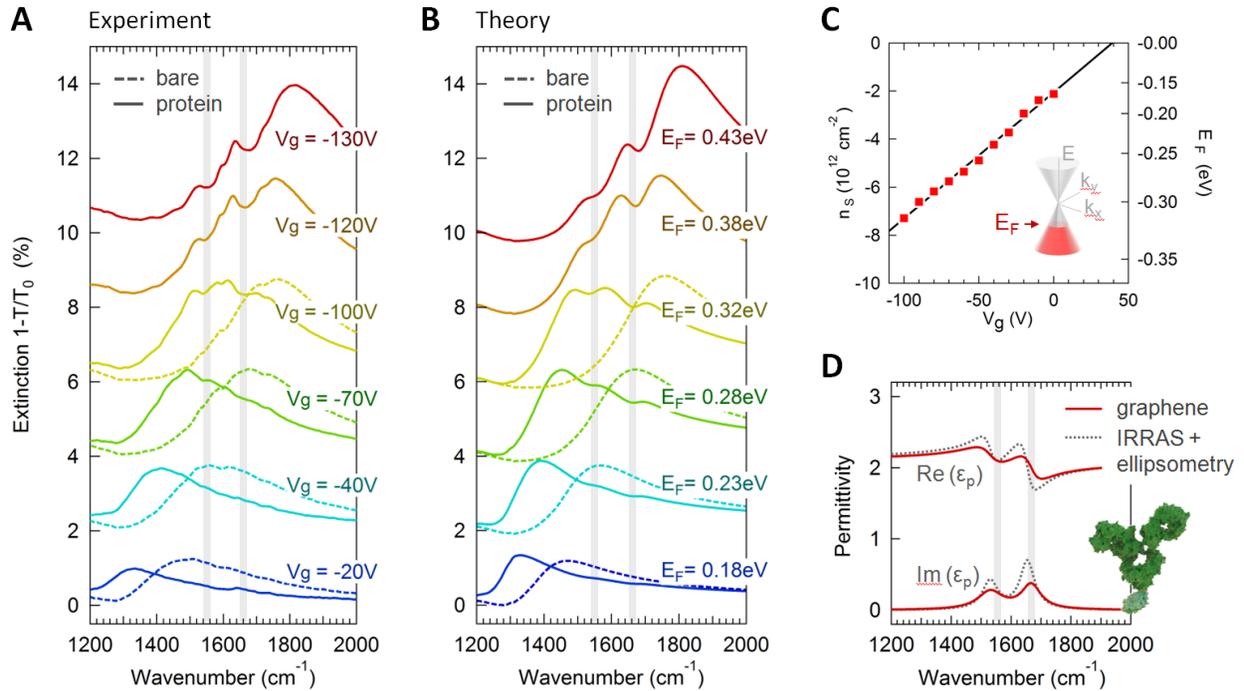

**Fig. 2. Mid-IR spectrum of the graphene biosensor.** (**A**) Extinction spectra of the graphene nanoribbon array ($W$=30nm, $P$=80nm) for bias voltages $V_g$ from -20V to -130V before (dashed curves) and after (solid curves) protein bilayer formation. Grey vertical strips indicate amide I and II vibrational bands of the protein. (**B**) Analytic calculation of the extinction spectra after fitting graphene and protein parameters to reproduce experimental data. (**C**) Graphene carrier density ($n_s$) and Fermi energy ($E_F$) extracted from experimental IR extinction spectra of the bare graphene nanoribbon array at different applied bias voltage $V_g$. (**D**) Permittivity of the protein bilayer extracted from the analytic fit to the experimental IR spectra (solid red curve) of the graphene biosensor compared to the permittivity extracted from IRRAS and ellipsometry measurements (dashed black curve).

**Fig. 3.**

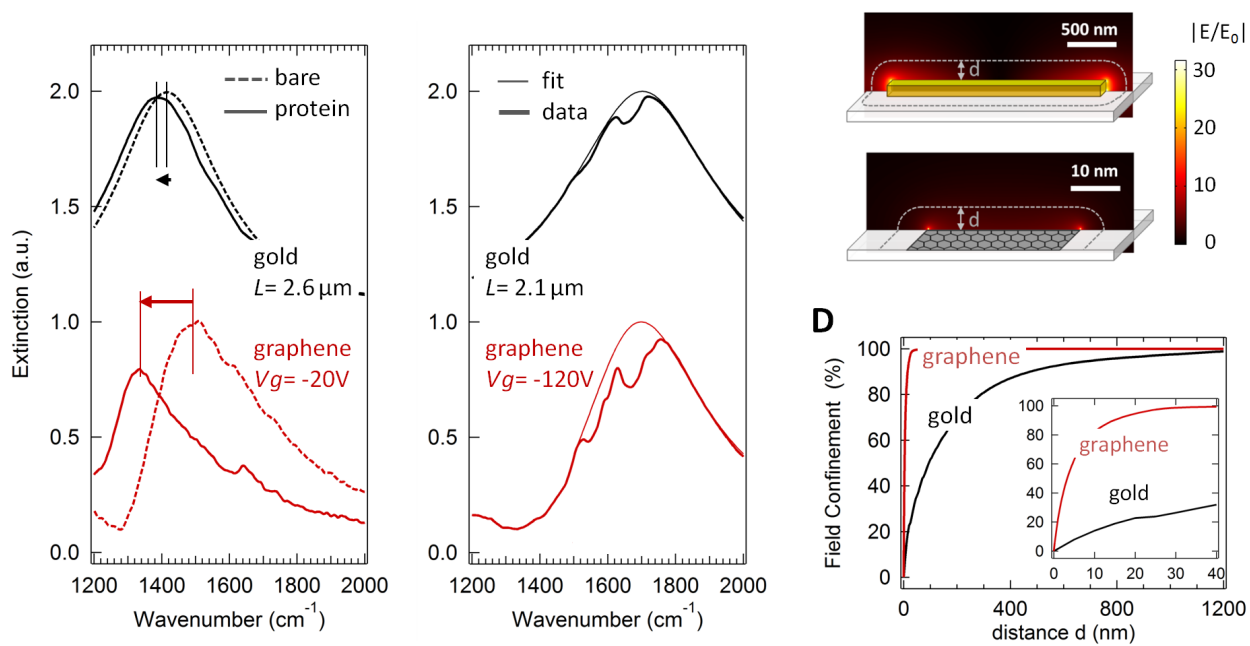

**Fig. 3. Graphene vs gold.** (**A**) Extinction spectra of graphene and gold nanoantenna arrays before (dashed curves) and after (solid curves) protein bilayer formation for plasmonic resonance peak away from the molecular vibration bands. The gold antennas have dimensions 2.6x0.2x0.1μm$^3$ while the graphene is biased to $V_g$=-20V. The spectral shift of the plasmonic resonance (indicated by horizontal arrows) shows the refractive index sensitivity of the biosensors. (**B**) Extinction spectra of graphene and gold biosensors after protein formation (thick curves) and fitting (thin curves) for plasmon peak overlapping with the molecular vibration bands. The gold antennas have dimensions 2.1x0.2x0.1μm$^3$ and the graphene gate voltage is $V_g$=-120V. The intensity of the spectral features at Amide I and II bands (1660-1550cm$^{-1}$) indicate the SEIRA sensitivity of the biosensors. (**C**) Near-field enhancement distribution $|E/E_0|$ in the plasmonic sensors operating at 1600cm$^{-1}$ resonance frequency. (**D**) Percentage of space-integrated near-field intensity confined within a volume extending a distance *d* outside the nanoantenna. Inset shows a zoom-in for *d* between 0 and 40nm.

# Supplementary Information

**Materials and Methods:**

**Nanofabrication of graphene nano-ribbon arrays.** Double-side polished float-zone silicon is used as substrate for the graphene optical device. A silicon dioxide layer of 280nm is grown on silicon through dry oxidation and the backside oxide is etched in $C_4F_8$ plasma. Graphene is grown by chemical vapor deposition on a copper catalyst and wet-transferred to the substrate. A 100nm layer of electron beam resist is spin coated on the chip. Nanoribbon arrays with widths W from 15nm to 100nm are exposed using 100keV electron beam lithography. After resist development graphene nanoribbons are etched in oxygen plasma at 300W for 12s. Resist stripping is done in acetone, followed by IPA and DI water rinsing.

**Nanofabrication of gold nano-antenna arrays.** Gold nanoantenna arrays are fabricated on a $CaF_2$ substrate. A MMA/PMMA resist layer is spin coated on the chip. Antenna arrays are exposed using 100keV electron beam lithography after the sputtering of a gold layer to avoid electron charging. The arrays are composed of dipoles with a width of 200nm, a length varying from 1.6μm to 2.2μm and a periodicity 25% larger than the dipole. The thin gold layer is wet etched followed by the photoresist development. Gold nanodipoles are formed by evaporation of Cr/Au 10-100nm and lift-off in acetone.

**Protein assay.** Protein A/G is diluted in phosphate-saline-buffer (PBS 1x) sterile solution at a 1mg/mL concentration. The sensor chip is incubated with protein A/G solution to allow protein physisorption. The chip is rinsed in PBST to remove unbound protein and agglomerates. Next, anti-Mouse IgG (Fc specific) antibody at a 0.5mg/mL concentration in PBS is incubated and followed by rinsing.

**Infrared spectroscopy measurements.** Extinction spectra are measured with a Fourier transform IR interferometer coupled to an IR microscope (Bruker Vertex 80V and Hyperion 3000). Transmission spectra are collected with linearly-polarized light through a Swcharzchild objective (NA=0.4, 15x) and measured by a liquid-nitrogen-cooled mercury-cadmium-tellurium detector. Extinction spectra ($1-T/T_0$) are calculated by normalizing the transmission spectrum by that of the chip without nanostructures ($T_0$). The chip is electrostatically biased during the measurements by applying a voltage ($V_g$ from 0V to -130V) between the silicon backside and a metallic pad connected to graphene.

Infrared reflection absorption (IRRAS) spectra are acquired for an 80º incidence angle, p-polarized light and a 2 mm aperture. IRRAS measurements are referenced to a bare gold mirror.

**Parameter extraction from analytic calculations.** The extraction of graphene parameters and protein permittivity are performed by adjusting the corresponding parameters in the analytic model to minimize the mean square error between experimental and calculated spectra. Firstly, the graphene Fermi level ($E_F$) for each voltage ($V_g$) and the graphene relaxation time ($\tau$) are adjusted so the spectra calculated for the corresponding graphene conductivity ($\sigma(\omega, \tau, EF)$) matched the experimental spectra of bare graphene nanoribbons. Secondly, a 2-Lorentzian protein permittivity model are automatically optimized to match the fingerprints detected for the highest bias voltages ($V_g$=-120V,-130V). The extracted values for graphene parameters are $\tau$=15fs and $E_F$=0.18eV to 0.43eV, and the protein permittivity parameters are: $n_\infty^2$=2.08, $\omega_1$=1668cm$^{-1}$, $\omega_2$=1532cm$^{-1}$, $\gamma_1$=78.1cm$^{-1}$, $\gamma_2$=101cm$^{-1}$, $S_1$=213cm$^{-1}$, $S_2$=200cm$^{-1}$.

**Electromagnetic simulations.** Electromagnetic simulations (Fig.S3) are conducted using a Finite Elements Method (HFSS) with periodic boundary conditions. Graphene is modeled as a two-dimensional surface with complex conductivity from Kubo formula *(24)* for the experimentally extracted graphene parameters. The protein bilayer is included as an 8 nm-thick layer having Lorentzian complex permittivity with parameters extracted from IRRAS measurements. Material parameters for Si, SiO$_2$, CaF$_2$ and Au are obtained from Palik *(22)*. Transmission and reflection spectra are normalized by the response of the bare device without graphene/gold nanostructures. Near-field distributions are computed by subtracting the incident and scattered plane waves from the total simulated field.

**Supplementary Text:**

**IRRAS characterization.** The expected normalized reflectance, for a thin film on thick substrate, can be derived from standard three-layer Fresnel equation and successive linear Taylor series approximation *(25)*.

$$\frac{R(d) - R(0)}{R(0)} \approx 2d \left(\frac{4\pi}{\lambda}\right) \left(\frac{\sin^2(\theta)}{\cos(\theta)}\right) Im\left(\frac{-1}{\epsilon}\right) \tag{S1}$$

where d and $\epsilon$ are the thickness and permittivity of the thin layer, respectively, and $\theta$ the incidence angle. For grazing incidence at $\theta = 80°$ it is obtained a 10-fold enhancement of the film absorption signal with respect to bare transmission.

To calculate the enhancement factor for graphene plasmonic biosensor, we performed IRRAS measurements of A/G-IgG protein bilayer on continuous graphene on gold (GR/Au) and gold only (Au) (Fig. S2). Prior to protein immobilization a ZEP photoresist layer was coated and stripped from the Au/GR chip to reproduce the same surface as in the graphene nanoribbon device.

We used Eq. S1 to fit IRRAS signal of A/G-IgG on Au to extract the protein bilayer permittivity $\epsilon$ under the following assumptions: *i*) d=8nm for the A/G-IgG bilayer as estimated by independent ellipsometric measurements, *ii*) the protein permittivity can be modelled as a Lorentz series with two oscillating terms accounting for amide-I and II bands.

(S2)

$$\epsilon(\omega) = n_\infty^2 + S_1{}^2/(\omega_1^2 - \omega^2 - j\omega\gamma_1) + S_2{}^2/(\omega_2^2 - \omega^2 - j\omega\gamma_2)$$

Assuming the non-dispersive component of the refractive index to be $n_\infty^2 = 2.10$, the extracted parameters for amide-I and amide-II are:
$\omega_1 = 1655 \text{cm}^{-1}$, $\omega_2 = 1530 \text{cm}^{-1}$, $\gamma_1 = 59.1 \text{cm}^{-1}$, $\gamma_2 = 61.6 \text{cm}^{-1}$, $S_1 = 258 \text{cm}^{-1}$, $S_2 = 194 \text{cm}^{-1}$.

**Fig. S1.**

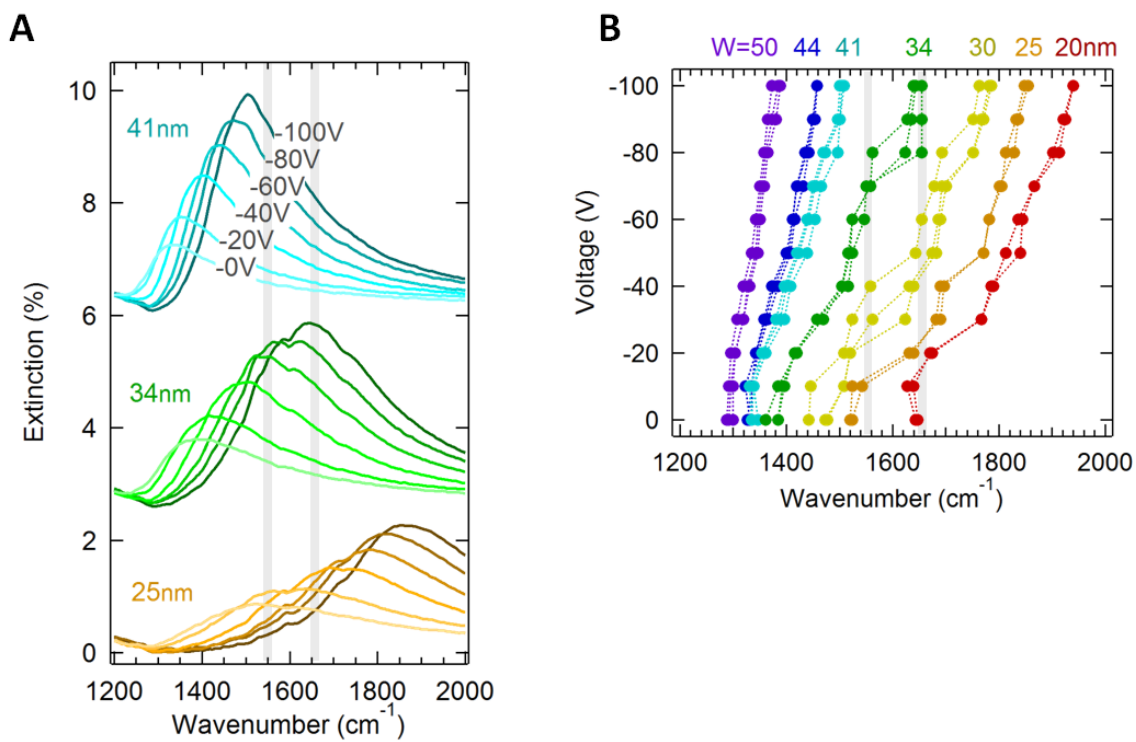

**Fig. S1. Mid-IR spectrum of bare graphene nanoribbon arrays. (A)** Measured extinction spectrums (1-T/$T_0$) of three different graphene nanoribbons arrays having widths W=41nm, 34nm, 25nm and for bias voltages Vg from 0V to -100V. The reference spectrum $T_0$ is the transmission through the bare substrate without graphene nanoribbons. The two dashed vertical lines indicate Amide I and II bands. **(B)** Measured resonance frequency of the fabricated graphene nanoribbon arrays for different W and Vg. Different replicas have been measured for each combination of W and Vg. Resonance frequency is extracted as the frequency of maximum extinction.

**Fig. S2.**

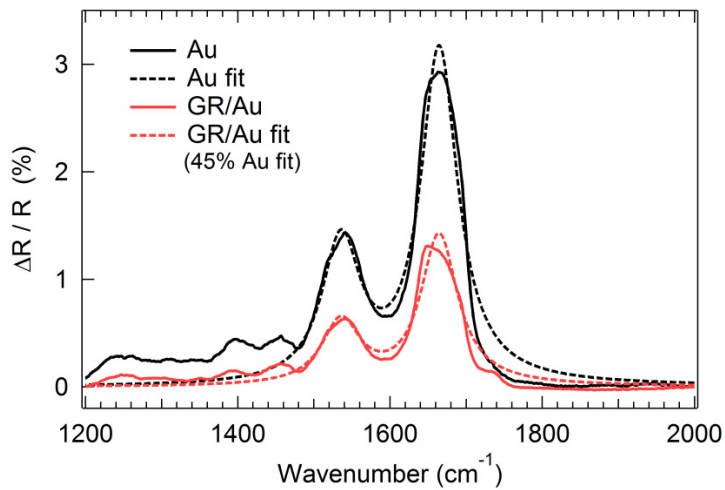

**Fig. S2. Protein characterization from IRRAS measurements.** (**A**) Infrared absorption (IRRAS) spectrum of the AG/IgG protein bilayer in bare gold and a graphene-coated gold mirror. The peaks at 1660cm$^{-1}$ and 1550cm$^{-1}$ correspond to Amide I and II bands, respectively. Solid lines represent measured absorption and dashed lines represent absorption fitting using a Lorentzian permittivity model. The protein mass absorbed by graphene is 45% of that absorbed by gold.

**Fig. S3.**

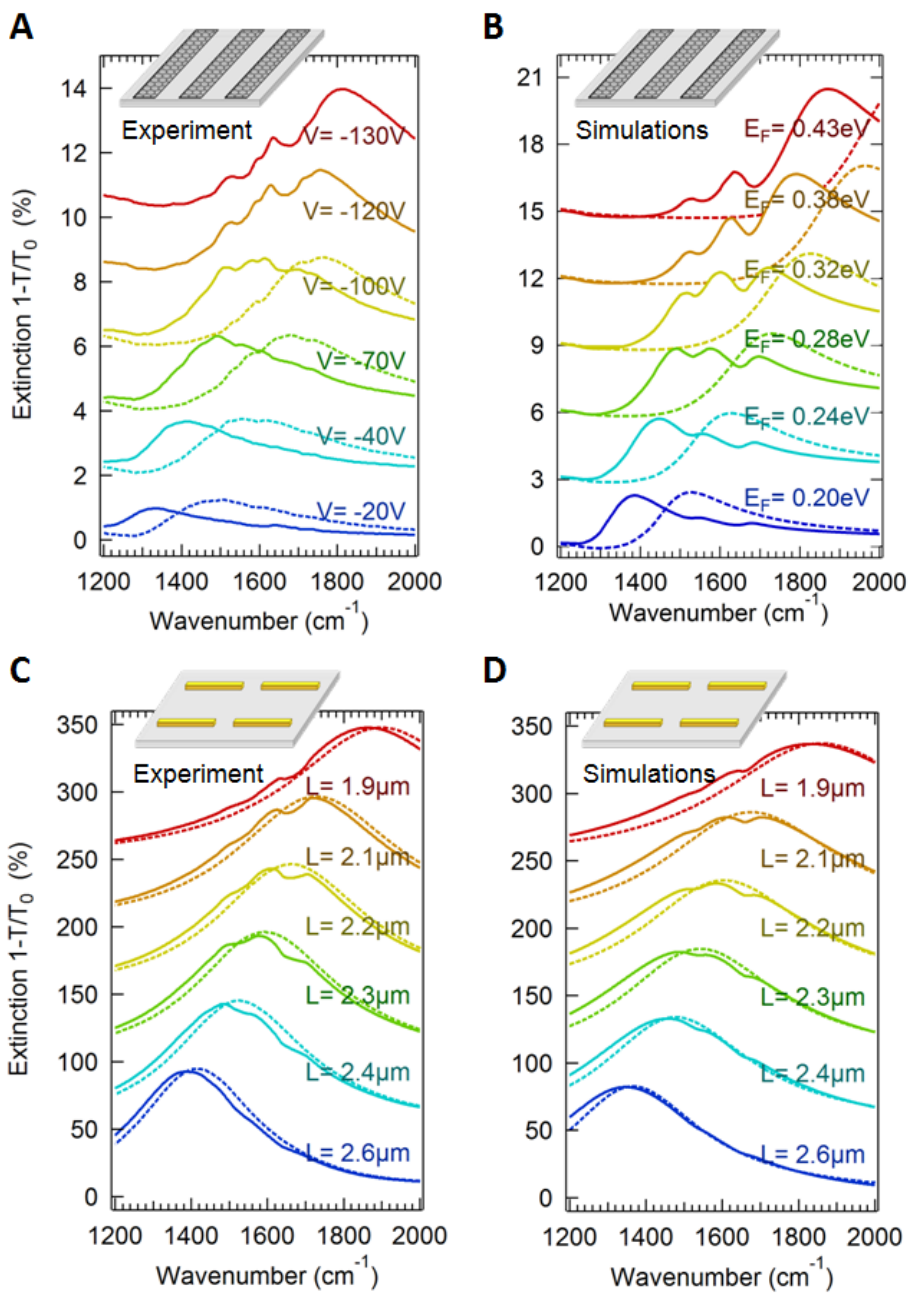

**Fig. S3. Experiment vs full-wave simulations.** Measured **(A)** and calculated **(B)** extinction spectrums for a graphene nanoribbon array with W=30nm, P=80nm and varying bias voltage and Fermi energy, respectively. Dashed lines represent the extinction of the bare graphene device and solid lines represent extinction after protein bilayer formation. Calculations are carried out using Finite Element Method. Measured **(C)** and calculated **(D)** extinction spectrums for a gold nanoantenna array with 200nm width, 100nm height and length varying length. Dashed and solid lines represent before and after protein bilayer formation.